\documentclass[amssymb,aps,12pt]{revtex4}

\usepackage{graphicx}
\usepackage{subfigure}

\newcommand{\be}{\begin{equation}}
\newcommand{\ee}{\end{equation}}
\newcommand{\bsmm}{B_s \rightarrow \mu^+ \mu^-}

\begin{document}
%\draft

\preprint{\begin{tabular}{l}
\hbox to\hsize{February, 2002 \hfill KAIST-TH 2002/12}\\[-3mm]
\hbox to\hsize{hep-ph/0202xxx \hfill KIAS-P02008}\\[5mm] \end{tabular} }

\title{
%Implications on SUSY breaking mediation mechanisms \\ from observing 
%$B_s \rightarrow \mu^+ \mu^-$ and the muon $(g-2)$ 
B-decays at Large $\tan\beta$ as a Probe of SUSY Breaking
\footnote{talk by S. Baek}
} 

\author{$^a$Seungwon Baek, ~$^b$P. Ko and ~$^b$Wan Young Song}
\affiliation{
$^a$ Korea Institute for Advanced Study, 207-43 Cheongryangri-dong, 
Seoul 130-012, 
 Korea \\}
\affiliation{
$^b$ Dep. of Physics, KAIST,  Daejeon 305-701, Korea \\}

%\date{\today}

%\maketitle
%\thispagestyle{empty}

\begin{abstract}
We consider $B_s \rightarrow \mu^+ \mu^-$ and the muon $(g-2)_\mu$ 
in various SUSY breaking mediation mechanisms. 
If the decay $B_s \rightarrow \mu^+ \mu^-$ is observed at Tevatron Run II 
with a branching ratio larger than $\sim 2 \times 10^{-8} $, the noscale 
supergravity (including the gaugino mediation), the gauge mediation scenario 
with small number of messenger fields and low messenger scale, 
and a class of anomaly mediation 
scenarios will be excluded, even if they can accommodate a large muon 
$(g-2)_\mu$. On the other hand, the minimal supergravity scenario and 
similar mechanisms derived from string models can accommodate this 
observation. 
\end{abstract}


\maketitle
%\narrowtext
%\tighten

%%%%%%%%%%%%%%%%%%%%%%%%%%%%%%%%%%%%%%%%%%%%%%%%%%%%%%%%%%%%%%%%%%%%%%%%%%%%%
%%%%%%%%%%%%%%%%%%%%%%%%%%%%%%%%%%%%%%%%%%%%%%%%%%%%%%%%%%%%%%%%%%%%%%%%%%%%%

%\section{Introduction}

The minimal supersymmetric standard model (MSSM) is one of the leading 
candidates for the physics beyond the standard model (SM). Its 
detailed phenomenology depends on soft SUSY breaking terms which contain
105 new parameters compared to the SM.
There are some interesting suggestions that have been put forward over 
the last two decades:  gravity mediation (SUGRA), gauge mediation (GMSB), 
anomaly mediation (AMSB), and gaugino mediation ($\tilde{g}$MSB), etc..  
Each mechanism predicts specific forms of soft SUSY breaking 
parameters at some messenger scale. 
It is most important to determine the soft parameters from various different
experiments, and compare the resulting soft SUSY breaking parameters with those
predicted in the aforementioned SUSY breaking mediation mechanisms. This 
process will provide invaluable informations on the origin of SUSY breaking,
which may be intrinsically rooted in very high energy regimes such as 
intermediate, GUT or Planck scales.

Direct productions of SUSY particles and measuring their properties are
indispensable for this purpose.
% However, the importance of indirect effects
%of SUSY particles through quantum loop corrections cannot be underestimated
%at all for the following reasons. First of all, the experimental errors in
%many low energy processes are now already (or will be in the near future) 
%at the level of probing the loop effects from SUSY particles. %: 
%If theoretical uncertainties mainly from QCD can be reduced at a 
%satisfactory level, we can study the indirect effect of SUSY particles 
%within various SUSY breaking mediation mechanisms. 
%Secondly, %the high energy processes available at colliders are usually 
%insensitive to the flavor structures of soft SUSY breaking parameters.
%On the other hand, 
%the low energy favor changing neutral current (FCNC) 
%processes such as $B - \overline{B} $ and $K - \overline{K}$ mixings as well
%as $B$ and $K$ decays are highly sensitive to %such 
%nontrivial flavor 
%structures in soft terms, in contrast to the most collider signals. 
%Thirdly, a certain region of parameter space 
%may not be accessible at Tevatron Run II, which is however sensitive to 
%$B_s \rightarrow \mu^+ \mu^-$. Therefore indirect search for SUSY effects 
%can be complementary to the direct search for SUSY particles. This was taken
%in recent works in the context of minimal SUGRA models, where it was found 
%that large positive $a_\mu^{\rm SUSY}$ implies a large enhancement of 
%$\bsmm$ branching ratio \cite{dedes}\cite{arnowitt}. 
However, indirect searches such as FCNC and/or CP violating processes,
can be complementary to the direct search.
%(Importance of Higgs 
%mediated $B_d^0 \rightarrow \mu^+ \mu^-$ was pointed out in Ref.~\cite{babu}.)

We considered the low energy processes $(g-2)_\mu$, 
%In this letter, we consider the low energy processes $(g-2)_\mu$, 
$B\rightarrow X_s \gamma$ and $B_s \rightarrow \mu^+ \mu^-$ for theoretically 
well motivated SUSY breaking mediation mechanisms~\cite{prl}: no scale scenario 
\cite{noscale} including $\tilde{g}$MSB \cite{ginomsb}, GMSB \cite{gmsb} 
and the minimal AMSB \cite{amsb} and some of variations 
\cite{gamsb,damsb,fi}. 
It turns out there are qualitative 
differences among some correlations for different SUSY breaking mediation
mechanisms~\cite{prl,tata}. 
Especially the branching ratio for $\bsmm$ turns out sensitive 
to the SUSY breaking mediation mechanisms, irrespective of the muon anomalous 
magnetic moment $a_\mu^{\rm SUSY}$ as long as $10 \times 10^{-10} 
\lesssim a_\mu^{\rm SUSY} \lesssim 40 \times 10^{-10}$. If $\bsmm$ is observed
at Tevatron Run II with a branching ratio larger than 
$\sim 2 \times 10^{-8}$, the GMSB with a small number of messenger fields 
with low messenger scale and a class of AMSB scenarios will be excluded. 
Only supergravity or GMSB with high messenger scale and large number of 
messenger fields and the deflected AMSB  would survive. 

%%%%%%%%%%%%%%%%%%%%%%%%%%%%%%%%%%%%%%%%%%%%%%%%%%%%%%%%%%%%%%%%%%%%%%%%%%%%%

%\section{Relevant processes and analysis procedures}

%%%%%%%%%%%%%%%%%%%%%%%%%%%%%%%%%%%%%%%%%%%%%%%%%%%%%%%%%%%%%%%%%%%%%%%%%%%%
%\subsection{Muon anomalous magnetic moment : $a_\mu$}
%%%%%%%%%%%%%%%%%%%%%%%%%%%%%%%%%%%%%%%%%%%%%%%%%%%%%%%%%%%%%%%%%%%%%%%%%%%%

The SUSY contributions to $a_\mu$ come from the chargino-sneutrino and 
the neutralino-smuon loop, the former of which is dominant in most parameter
space. 
%Schematically, the result can be written as \cite{Martin:2001st}
%\begin{equation}
%a_\mu^{\rm SUSY} = 
%{\tan\beta \over 48 \pi}{m_\mu^2 \over M_{\rm SUSY}^2}
%( 5 \alpha_2 + \alpha_1 ) 
%\end{equation}
%in the limit where all the superparticles have the same mass $M_{\rm SUSY}$.
In particular, $\mu > 0$ implies $a_\mu^{\rm SUSY} > 0$ in our convention. 
The deviation between the new BNL data 
\cite{g-2exp} and the most recently updated SM prediction\cite{g-2th} 
based on the $\sigma ( e^+ e^- \rightarrow $ hadrons) data  is 
$( 33.9 \pm 11.2) \times 10^{-10}$.%, which is about $3 \sigma$ deviation.
%can not be taken as a strong indication for new physics beyond the SM. 
On the other hand, the deviation becomes smaller if the hadronic tau decays 
are used. Therefore, we do not use $a_\mu$ as a constraint except for 
$a_\mu >0$, and give predictions for it in this letter.

%If the data is updated with smaller statistical 
%and systematic errors and theoretical uncertainties, $a_\mu^{\rm SUSY}$ 
%could provide a useful constraint on SUSY parameter space. 

%%%%%%%%%%%%%%%%%%%%%%%%%%%%%%%%%%%%%%%%%%%%%%%%%%%%%%%%%%%%%%%%%%%%%%%%%%%%
%\subsection{$B \rightarrow X_s \gamma$ and $B \rightarrow X_s l^+ l^-$}
%%%%%%%%%%%%%%%%%%%%%%%%%%%%%%%%%%%%%%%%%%%%%%%%%%%%%%%%%%%%%%%%%%%%%%%%%%%%

It has long been known that the $B \rightarrow X_s \gamma$ branching ratio 
puts a severe constraint on many new physics scenarios including weak scale 
SUSY models. The magnetic dipole coefficient $C_{7\gamma}$ for this decay 
gets contributions from SM, charged Higgs and SUSY particles in the loop. 
The charged Higgs contributions always add up to the SM contributions, 
thereby increasing the rate. On the other hand, the last (mainly by the stop 
- chargino loop) can interfere with the SM and the charged Higgs contributions 
either in a constructive or destructive manner depending on the sign of 
$\mu M_{\tilde{g}}$.
%Since the SM prediction is in good agreement with the data~\cite{bsg}, %: 
%$B ( B \rightarrow X_s \gamma )_{\rm exp} = ( 3.21 \pm 0.43_{(stat)} 
%\pm 0.27_{(sys) -0.10(th)}^{~~~~~~~+0.18} ) \times 10^{-4}$~\cite{bsg}, 
%from a 5.8 fb$^{-1}$ data sample~\cite{bsg}, 
%there is very little room for new physics contributions.  
Note that the 
positive $a_\mu^{\rm SUSY} $ picks up $\mu > 0$ (for $M_2 >0$)
in our convention.
Fortunately, this results in destructive interference of
the stop-chargino loop with the SM and the charged 
Higgs contribution in $B\rightarrow X_s \gamma$ decay, in all the models
considered except the AMSB scenario.
%In turn, this prefers 
%a positive $\mu M_{\tilde{g}}$ in many SUSY breaking scenarios except for
%the AMSB scenario in which $\mu M_{\tilde{g}} < 0$ \cite{Feng:1999hg}. 
In the AMSB scenario, the constructive interference between the 
stop-chargino loop and the SM contributions to $B\rightarrow X_s \gamma$, 
increases the rate even more. Therefore the AMSB scenario is 
strongly constrained if $a_\mu^{\rm SUSY} > 0$.  

%%%%%%%%%%%%%%%%%%%%%%%%%%%%%%%%%%%%%%%%%%%%%%%%%%%%%%%%%%%%%%%%%%%%%%%%%%%%
%\subsection{Hall--Rattazzi--Sarid Effect}
%%%%%%%%%%%%%%%%%%%%%%%%%%%%%%%%%%%%%%%%%%%%%%%%%%%%%%%%%%%%%%%%%%%%%%%%%%%%

Another important effect is the nonholomorphic SUSY QCD corrections to the 
$h b \bar{b}$ couplings in the large $\tan\beta$ limit: the 
Hall-Rattazzi-Sarid (HRS) effect \cite{hall}. Also, the stop - chargino 
loop could be quite important for large $A_t$ and $y_t$ couplings. One can 
summarize these effects as the following relation between the bottom quark 
mass and the bottom Yukawa coupling $y_b$:
\begin{equation}
m_b = y_b {\sqrt{2} M_W \cos\beta \over g}~( 1 + \Delta_b )
\end{equation} 
where the explicit form of $\Delta_b$ can be found in Ref.~\cite{logan}.
In the large $\tan\beta$ limit, the SUSY loop correction $\Delta_b$ which is 
proportional to $\mu M_{\tilde{g}} \tan\beta$ can be large as well with 
either sign, depending on the signs of the $\mu$ parameter and the gluino 
mass parameter $M_{\tilde{g}}$. In particular, the bottom Yukawa coupling 
$y_b$ becomes too large and nonperturbative, when $\mu > 0$ in the AMSB 
scenario, since the sign of $\Delta_b$ would be negative. This puts 
additional constraint on $\tan\beta \lesssim 35$ for the positive $\mu$ in 
the AMSB scenario. 
%Other related issues with Higgs mediated flavor changing 
%neutral current processes in SUSY models at large $\tan\beta$ can be found
%in Refs.~\cite{Hamzaoui:1998nu}.

%%%%%%%%%%%%%%%%%%%%%%%%%%%%%%%%%%%%%%%%%%%%%%%%%%%%%%%%%%%%%%%%%%%%%%%%%%%
%\subsection{$B_s \rightarrow \mu^+ \mu^-$} 
%%%%%%%%%%%%%%%%%%%%%%%%%%%%%%%%%%%%%%%%%%%%%%%%%%%%%%%%%%%%%%%%%%%%%%%%%%%

The decay $B_s \rightarrow \mu^+ \mu^-$ has a very small branching ratio  
in the SM ($(3.7 \pm 1.2) \times 10^{-9}$)\cite{Buchalla:1995vs}. 
But it can occur with much higher 
branching ratio in SUSY models when $\tan\beta$ is large, because the Higgs 
exchange contributions can be significant for large $\tan\beta$
\cite{Hamzaoui:1998nu}\cite{dedes}. 
The branching ratio for $B_s \rightarrow \mu^+ \mu^-$ is proportional to 
$\tan^6 \beta$ for large $\tan\beta$.
%The HRS effect can further modify the result in either 
%direction depending on the sign$(\mu)$. For $\mu > 0$, the enhancement 
%becomes less pronounced due to this HRS effect. 
Thus this decay may be observable 
at the Tevatron Run II down to the level of $2 \times 10^{-8}$, and could be
complementary to the direct search for SUSY particles at the Tevatron Run II
in the large $\tan\beta$ region.%  as discussed in Ref.~\cite{arnowitt}.
%as discussed in Ref.~\cite{arnowitt}.
%. Therefore 
%this decay can cover parameter space (large $\tan\beta$ region) which is not 
%accessible by direct search for SUSY particles at the Tevatron Run II, as
%discussed in Ref.~\cite{arnowitt}.

%%%%%%%%%%%%%%%%%%%%%%%%%%%%%%%%%%%%%%%%%%%%%%%%%%%%%%%%%%%%%%%%%%%%%%%%%%%%%
%\subsection{ Constraints }
%%%%%%%%%%%%%%%%%%%%%%%%%%%%%%%%%%%%%%%%%%%%%%%%%%%%%%%%%%%%%%%%%%%%%%%%%%%%%

In the following, we consider three aforementioned SUSY breaking mediation 
mechanisms. Each scenario gives definite predictions for the soft terms 
at some messenger scale. 
We use  renormalization group equations  in order to get soft parameters 
at the electroweak scale, impose the radiative electroweak symmetry breaking 
(REWSB) condition and then obtain particle spectra and mixing angles. 
Then we impose the direct search limits on Higgs and SUSY particles~\cite{prl}. 
%The most stringent limits turns out to be the neutral Higgs mass 
%bound ($m_h^{\rm SM} >113.5$ GeV) and $m_{\tilde{\tau} } > 71~{\rm GeV}$. 
%For the GMSB scenario, the LSP 
%is always very light gravitinos, and we impose 
%$m_{\rm NLSP}^{\rm GMSB} > 100 ~{\rm GeV}$, which is stronger than other 
%experimental bounds on SUSY particle masses. In order to be as model 
%independent as possible, we do not assume that the LSP is color and/or 
%charge neutral (except for the GMSB scenario where the gravitino is the LSP), 
%nor do we impose the color-charge breaking minima or the unbounded from below 
%constraints, since one can always find ways out. 
Also we impose the $B\rightarrow X_s \gamma$ branching ratio as a constraint 
with a conservative bound (at 95 \% C.L.) considering theoretical 
uncertainties related with QCD corrections: 
$2.0 \times 10^{-4} < B(B\rightarrow X_s \gamma) < 4.5 \times 10^{-4}$
\cite{bsg}. 

The correlation between $a_\mu^{\rm SUSY}$ and $\bsmm$ were recently studied
in the minimal SUGRA scenario~\cite{dedes}\cite{arnowitt}. 
The result is  that the positive large 
$a_\mu^{\rm SUSY}$ implies that $B( \bsmm )$ can be enhanced by a few orders 
of magnitude compared to the SM prediction, and can be reached at the 
Tevatron Run II. The $\tilde{g}$MSB scenario, which finds a natural setting 
in the brane world scenarios, leads to the no-scale SUGRA type boundary 
condition for soft parameters, in which scalar mass and trilinear scalar 
terms all vanish at GUT scale, $B = m_{ij}^2 = A_{ijk} = 0$ and only gaugino 
masses are non-vanishing. The result is shown in Fig.~1(a).
Assuming the gaugino mass unification at GUT scale, 
we find that overall phenomenology of $\tilde{g}$MSB scenario 
(and the noscale scenario) in the $a_{\mu}^{\rm SUSY}$ and 
$B_s \rightarrow \mu^+ \mu^-$ is similar to the mSUGRA scenario 
(see Ref.~\cite{bks} for details including $B\rightarrow X_s l^+ l^-$). 
In the allowed parameter space, the $a_\mu^{\rm SUSY}$ can easily become 
upto $\sim 60 \times 10^{-10}$. But the branching ratio for $\bsmm$ is 
always smaller than $2 \times 10^{-8}$ and becomes unobservable at the 
Tevatron Run II. %, as in the GMSB or AMSB models (see below). 
The reason is that the large $\tan\beta$ region, 
where the branching ratio for $\bsmm$ can be much
enhanced, is significantly constrained by stau or smuon mass bounds and 
the lower bound of $B\rightarrow X_s \gamma$. 
Therefore if the $a_\mu^{\rm SUSY}$ turns out to be positive and the decay 
$\bsmm$ is observed at the Tevatron Run II, the $\tilde{g}$MSB scenario 
would be excluded.

\begin{figure}[thb]
\centering
\subfigure[]{
\includegraphics[width=7cm,height=5cm]{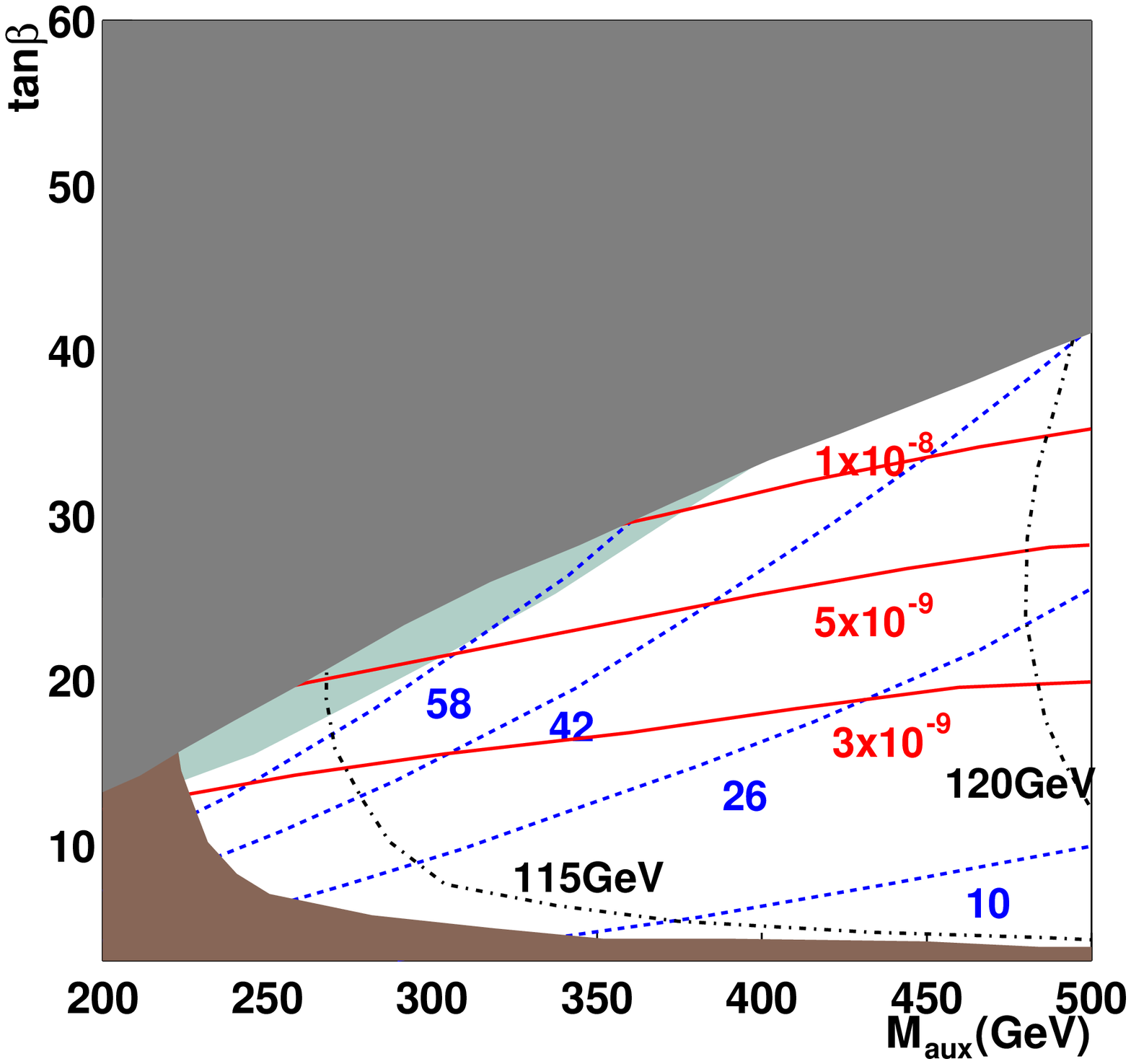}}
\subfigure[]{
\includegraphics[width=7cm,height=5cm]{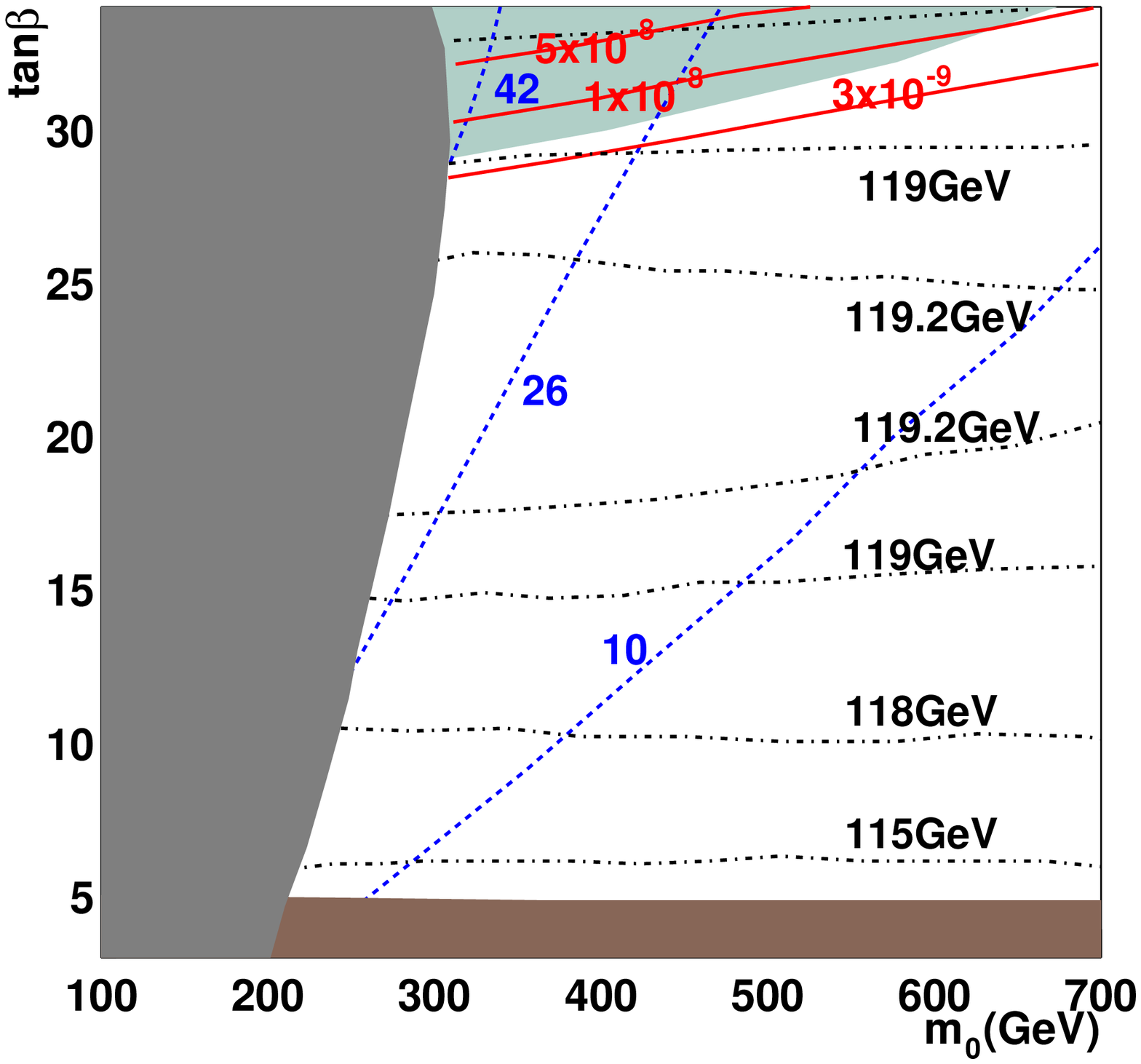}
\label{fig:amsb1}
}
%\label{fig:gmsb1}
%\includegraphics[width=7cm,height=5cm]{gmsb-1-6b.ps}}
%\centerline{\epsfxsize=7cm \epsfbox{gmsb-1-6b.ps}}
%\begin{picture}(150,150)(0,0)
%\put(0,220){
%\begin{rotate}{-90}
%\centerline{\epsfxsize=5.3cm 
%\epsfbox{figs/gmsb-1-6/tb-M1-1-6.ps}
%}
%\end{rotate}
%}
%\end{picture}
\caption{The contour plots for $a_\mu^{\rm SUSY}$ in unit of $10^{-10}$ 
(in the blue short dashed curves), the lightest neutral Higgs mass (in the 
black dash-dotted curves) and the Br ($B_s \rightarrow \mu^+ \mu^- $) 
(in the red solid curves) for (a) the $\tilde{g}$MSB scenario,
%(b) the the GMSB scenario in the 
%$( M_1, \tan\beta)$ plane with $N = 1$ and $M = 10^6$ GeV. 
%(b) the AMSB scenario in the $( m_0, \tan\beta)$ plane for $M_{\rm aux} = 50$ TeV.
(b) the AMSB scenario for $M_{\rm aux} = 50$ TeV.
The dark regions are excluded by the bounds from direct searches.
} 
\end{figure}

The `pure' AMSB model has the tachyonic slepton problem. For phenomenological
study we take the `minimal' AMSB model which has additional 
universal scalar mass $m_0^2$
at the GUT scale~\cite{Feng:1999hg}. 
It is specified by the following four parameters :
$\tan\beta, ~{\rm sign}(\mu), ~m_0,~ M_{aux}$.
In Fig.~\ref{fig:amsb1}, we show the contour plots for the $a_\mu^{\rm SUSY}$ 
and $B ( B_s \rightarrow \mu^+ \mu^- )$ in the $( m_0 , \tan\beta)$ plane for
$M_{\rm aux} = 50$ TeV. 
%The low $\tan\beta$ region is excluded by the lower
%limit on the neutral Higgs boson, and the small $m_0$ region is excluded by
%the LSP becomes staus and is strongly constrained by 
%the stau mass bound (the light dark region). 
In the case of the AMSB scenario with $\mu > 0$, 
the $B\rightarrow X_s \gamma$ constraint is even stronger compared
to other scenarios %(the shaded region in Fig.~\ref{fig:amsb1}). 
and  almost all the parameter space with large 
$\tan\beta > 30$ is excluded. %by the $B\rightarrow X_s \gamma$. 
%Also stops are relatively heavy in this scenario mainly due to the 
%universal addition of $m_0^2$. 
Therefore the branching ratio for $\bsmm$ is smaller than 
$4 \times 10^{-9}$, and this process becomes unobservable at the Tevatron 
Run II.  If the decay $\bsmm$ is observed at the Tevatron Run II, 
the minimal AMSB scenario would be excluded. 

\begin{figure}[thb]
\centering
\subfigure[]{
\includegraphics[width=7cm,height=5cm]{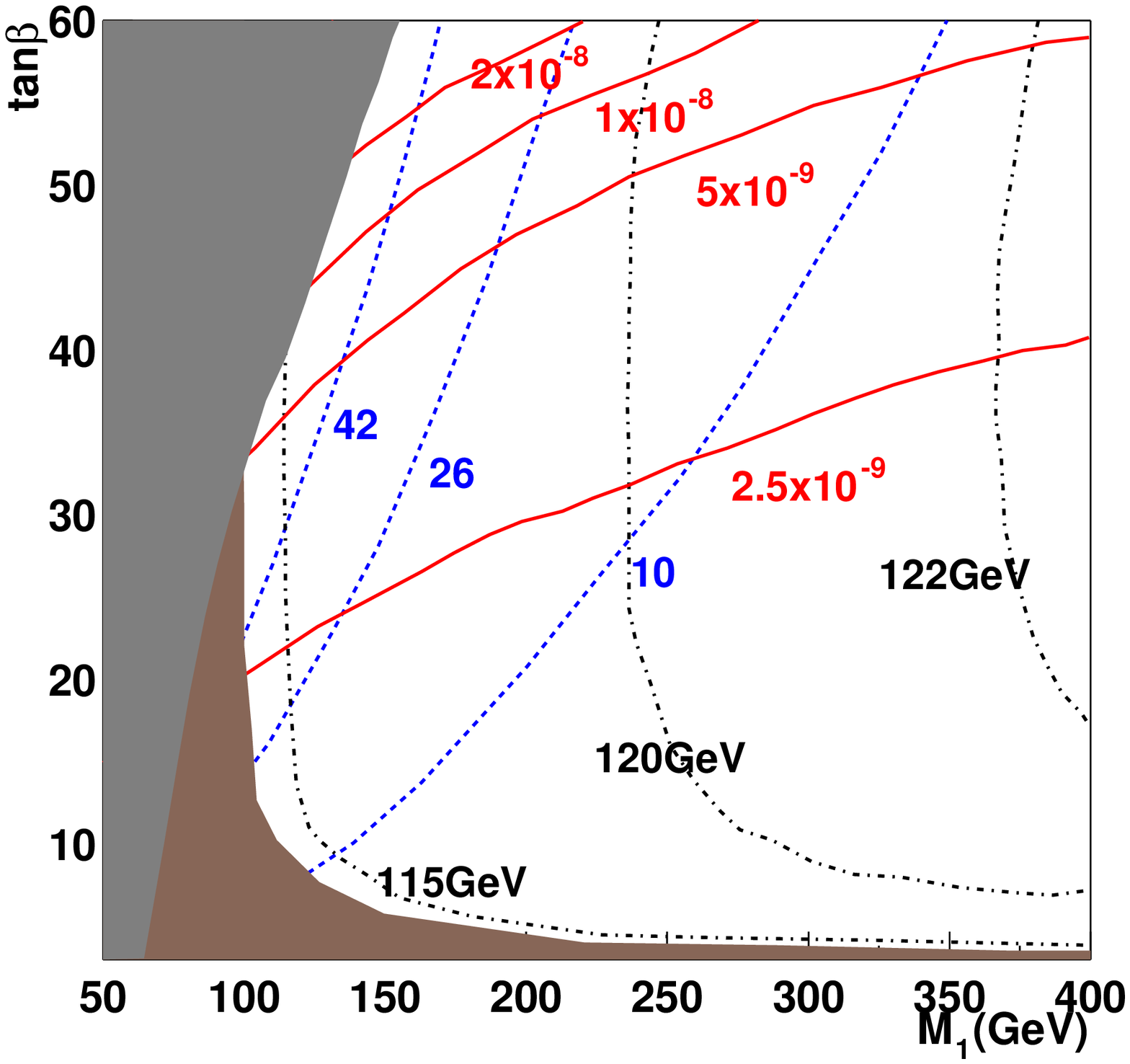}
\label{fig:gmsb1}
}
\subfigure[]{
\includegraphics[width=7cm,height=5cm]{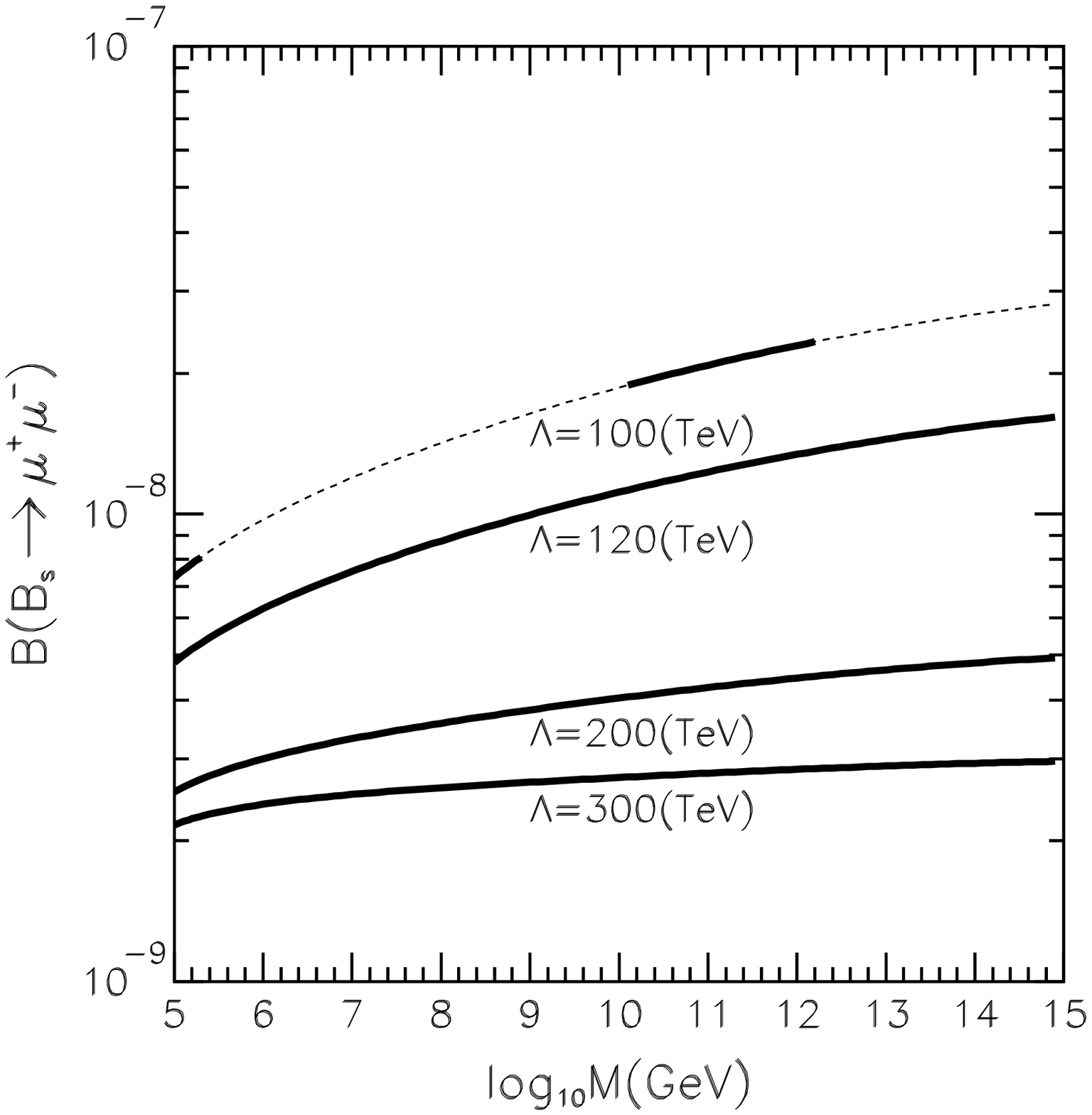}
\label{fig:gmsb2}
}
%\centerline{\epsfxsize=7cm \epsfbox{brbs-m-log.ps}}
%\begin{picture}(150,150)(0,0)
%\put(0,220){
%\begin{rotate}{-90}
%\centerline{\epsfxsize=5.3cm 
%\epsfbox{figs/gmsb-1-6/tb-M1-1-6.ps}
%}
%\end{rotate}
%}
%\end{picture}
\caption{%The correlation between the 
(a) The contour plots for the $a_\mu^{\rm SUSY}$,
$m_{h^0}$, and $B(\bsmm)$ with $N=1$ and $M=10^6$ GeV.
(b) The branching ratio for 
$B_s \rightarrow \mu^+ \mu^-$ as a function of the messenger scale $M$ 
in the GMSB with $N=1$ for various $\Lambda$'s with a fixed  
$\tan\beta = 50$. The dashed parts are excluded by the direct search limits
on the Higgs and SUSY particle masses.}
%: $3-20$ in black, $20-30$ in red, $30-40$ 
%in green and $40-50$ in blue.} % and $50-60$ in violet.}
\end{figure}

GMSB scenarios are specified by the following set of parameters: 
$M$, $N$, $\Lambda$, $\tan\beta$ and sign($\mu$), where $N$ is the number of 
messenger superfields, $M$ is the messenger scale, and the $\Lambda$ is SUSY 
breaking scale, $\Lambda \approx \langle F_X \rangle / \langle X \rangle$.
In Fig.~\ref{fig:gmsb1}, we show the contour plots for the $a_\mu^{\rm SUSY}$,
$m_{h^0}$, and $B(\bsmm)$ with $N=1$ and $M=10^6$ GeV.
%and $B ( B_s \rightarrow \mu^+ \mu^- )$ in the $( M_1 , \tan\beta)$ plane for
%$N_{\rm mess} = 1$ and $M_{\rm mess} = 10^6$ GeV, where the parameter 
%$\Lambda$ has been traded into the bino mass parameter $M_1$.
%The left dark region is 
%excluded by direct search limits on Higgs boson masses, and the gray region 
%is excluded by the limit on the NLSP mass.
%Since the messenger scale is low, 
%the SUSY flavor problem is evaded in the GMSB scenarios. The 
%the flavor changing amplitude involving the 
%gluino-squark is negligible and only the chargino-upsquark contribution is
%important to $B\rightarrow X_s \gamma$.  Also, in the GMSB scenario with low 
For low messenger scale, the charged Higgs and stops are heavy and their effects on 
the $B\rightarrow X_s \gamma$ and $\bsmm$ are small. And the $A_t$ is small
since it can generated by only RG running, so that the stop mixing angle 
becomes small. These effects lead to very small branching ratio for $\bsmm$
($\lesssim 10^{-8}$), %and 
making this decay unobservable at the Tevatron 
Run II.  On the other hand, the $a_\mu^{\rm SUSY}$ can be as large as 
$60 \times 10^{-10}$. For a given $N$, $B(\bsmm)$ increases as $M$ due to
RG effect (see Fig.~\ref{fig:gmsb2}). Also for larger $N$, $B(\bsmm)$ is 
enhanced because the scalar masses are suppressed relative to the gaugino 
masses.

%This general feature of the minimal AMSB scenario is still valid 
%in the gaugino assisted AMSB scenario 
%%, which is a hybrid of the $\tilde{g}$MSB and $AMSB$
%\cite{gamsb},
%where the scalar mass terms receive gauge-charge dependent positive 
%contributions from the MSSM gauge multiplets living 
%in the bulk, in addition to the pure anomaly mediation term.
%%Thus the overall phenomenology is the similar to the previous pure AMSB
%%with universal $m_0^2$, and we find $B( B_s \rightarrow \mu^+ \mu^-)$ 
%%never gets larger than $2 \times 10^{-8}$, which is not accessible at
%%the Tevatron. 
%On the other hand, in the deflected AMSB scenario \cite{damsb}, the soft 
%SUSY breaking parameters are shifted from the pure AMSB case when heavy
%particles are integrated out and the tachyonic slepton problem is solved.
%Also the gluino mass parameter can flip the sign when the number of gauge
%charged messengers are increased. In this case the $B\rightarrow X_s \gamma$
%constraint becomes weaker %as in the mSUGRA or GMSB scenarios. Overall 
%and overall phenomenology 
%% in the $a_\mu$ and $B_s \rightarrow \mu^+ \mu^-$ are 
%is similar to the mSUGRA case (see Ref.~\cite{bks} for more details).
%In case the Fayet-Iliopoulos term is employed to cure the tachyonic 
%slepton problem, the allowed $\tan\beta$ is rather small \cite{fi} so that 
%the branching ratio for $B_s \rightarrow \mu^+ \mu^-$ cannot be enhanced 
%to be observed at the Tevatron Run II.

In conclusion, we showed that there are qualitative differences in 
correlations among $(g-2)_{\mu}$, $B\rightarrow X_s \gamma$, 
and $B_s \rightarrow \mu^+ \mu^-$ in various models for SUSY breaking 
mediation mechanisms, even if all of them can accommodate the muon $a_\mu$:
$10\times 10^{-10} \lesssim a_\mu^{\rm SUSY} \lesssim 40 \times 10^{-10}$. 
Especially, if the $\bsmm$ decay is observed at Tevatron Run II with 
the branching ratio greater than $2 \times 10^{-8}$, the GMSB 
with low number of messenger fields $N$ and %the minimal and gaugino assisted 
certain class of AMSB scenarios would be excluded. 
On the other hand, the minimal supergravity scenario and similar mechanisms 
derived from string models and the deflected AMSB scenario can accommodate 
this observation \cite{bks} without difficulty for large $\tan\beta$. 
Therefore search for $\bsmm$ decay at the Tevatron Run II would provide 
us with important informations on the SUSY breaking mediation mechanisms,
independent of informations from direct search for SUSY particles at high 
energy colliders. 

%%%%%%%%%%%%%%%%%%%%%%%%%%%%%%%%%%%%%%%%%%%%%%%%%%%%%%%%%%%%%%%%%%%%%%%%%%

\acknowledgments
This work is supported in part by BK21 Haeksim program and also 
by KOSEF SRC program through CHEP at Kyungpook National University.

%\vspace{-1.5cm}

%\vspace{-1.5cm}

\vfil\eject

\begin{thebibliography}{99}

\bibitem{prl}
%\cite{Baek:2002rt}
%\bibitem{Baek:2002rt}
S.~w.~Baek, P.~Ko and W.~Y.~Song,
%``Implications on SUSY breaking mediation mechanisms from observing  B/s $\to$ mu+ mu- and the muon (g-2),''
arXiv:hep-ph/0205259 (to appear in Phys. Rev. Lett.).
%%CITATION = HEP-PH 0205259;%%
\bibitem{noscale}
J.~R.~Ellis, C.~Kounnas and D.~V.~Nanopoulos,
%``No Scale Supersymmetric Guts,''
Nucl.\ Phys.\ B {\bf 247}, 373 (1984).
%%CITATION = NUPHA,B247,373;%%

\bibitem{ginomsb} 
Z.~Chacko, M.~A.~Luty and E.~Ponton,
%``Massive higher-dimensional gauge fields as messengers of supersymmetry  
%breaking,''
JHEP {\bf 0007}, 036 (2000);
%[arXiv:hep-ph/9909248];
%%CITATION = HEP-PH 9909248;%%
D.~E.~Kaplan, G.~D.~Kribs and M.~Schmaltz,
%``Supersymmetry breaking through transparent extra dimensions,''
Phys.\ Rev.\ D {\bf 62}, 035010 (2000);
%[arXiv:hep-ph/9911293];
%%CITATION = HEP-PH 9911293;%%
Z.~Chacko, M.~A.~Luty, A.~E.~Nelson and E.~Ponton,
%``Gaugino mediated supersymmetry breaking,''
JHEP {\bf 0001}, 003 (2000).
%[arXiv:hep-ph/9911323].
%%CITATION = HEP-PH 9911323;%%

\bibitem{gmsb} For a recent review, see G.~F.~Giudice and R.~Rattazzi,
%``Theories with gauge-mediated supersymmetry breaking,''
Phys.\ Rept.\  {\bf 322}, 419 (1999).
%[arXiv:hep-ph/9801271].
%%CITATION = HEP-PH 9801271;%%

\bibitem{amsb} 
L.~Randall and R.~Sundrum,
%``Out of this world supersymmetry breaking,''
Nucl.\ Phys.\ B {\bf 557}, 79 (1999);
%[arXiv:hep-th/9810155];
%%CITATION = HEP-TH 9810155;%%
G.~F.~Giudice, M.~A.~Luty, H.~Murayama and R.~Rattazzi,
%``Gaugino mass without singlets,''
JHEP {\bf 9812}, 027 (1998);
%[arXiv:hep-ph/9810442];
%%CITATION = HEP-PH 9810442;%%
T.~Gherghetta, G.~F.~Giudice and J.~D.~Wells,
%``Phenomenological consequences of supersymmetry with anomaly-induced  
%masses,''
Nucl.\ Phys.\ B {\bf 559}, 27 (1999).
%[arXiv:hep-ph/9904378].
%%CITATION = HEP-PH 9904378;%%

\bibitem{gamsb} D.~E.~Kaplan and G.~D.~Kribs, JHEP {\bf 0009}, 048 (2000).

\bibitem{damsb} A.~Pomarol and R.~Rattazzi, JHEP {\bf 9905}, 013 (1999);
R.~Rattazzi, A.~Strumia, J.~D.~Wells, Nucl.~Phys. {\bf B576}, 3 (2000).

\bibitem{fi} 
I.~Jack and D.~R.~Jones,
%``Fayet-Iliopoulos D-terms and anomaly mediated supersymmetry breaking,''
Phys.\ Lett.\ B {\bf 482}, 167 (2000);
%[arXiv:hep-ph/0003081];
%%CITATION = HEP-PH 0003081;%%
N.~Arkani-Hamed, D.~E.~Kaplan, H.~Murayama and Y.~Nomura,
%``Viable ultraviolet-insensitive supersymmetry breaking,''
JHEP {\bf 0102}, 041 (2001)
%[arXiv:hep-ph/0012103].
%%CITATION = HEP-PH 0012103;%%

%\cite{Mizukoshi:2002gs}
%\bibitem{Mizukoshi:2002gs}
\bibitem{tata}
J.~K.~Mizukoshi, X.~Tata and Y.~Wang,
%``Higgs-mediated leptonic decays of B/s and B/d mesons as probes of  supersymmetry,''
arXiv:hep-ph/0208078.
%%CITATION = HEP-PH 0208078;%%


\bibitem{g-2exp}
G.~W.~Bennett {\it et al.}  [Muon g-2 Collaboration],
%``Measurement of the positive muon anomalous magnetic moment to 0.7 ppm,''
Phys.\ Rev.\ Lett.\  {\bf 89}, 101804 (2002)
[Erratum-ibid.\  {\bf 89}, 129903 (2002)].
%[arXiv:hep-ex/0208001].
%%CITATION = HEP-EX 0208001;%%

\bibitem{g-2th}
M.~Davier {\it et al.}, %, S.~Eidelman, A.~Hocker and Z.~Zhang,
%``Confronting spectral functions from e+ e- annihilation and tau decays:  Consequences for the muon magnetic moment,''
arXiv:hep-ph/0208177.
%%CITATION = HEP-PH 0208177;%%

%\bibitem{g-2}
%M.~Knecht and A.~Nyffeler, Phys.\ Rev.\ D {\bf 65}, 073034 (2002);
%M.~Knecht, A.~Nyffeler, M.~Perrottet, and E.~de Rafael,
%Phys.\ Rev.\ Lett.\  {\bf 88}, 071802 (2002);
%M.~Hayakawa and T.~Kinoshita, arXiv:hep-ph/0112102;
%A.~Czarnecki, and K.~Melnikov, Phys.\ Rev.\ Lett.\  {\bf 88}, 071803 (2002).

\bibitem{hall} 
L.~J.~Hall, R.~Rattazzi and U.~Sarid,
%``The Top quark mass in supersymmetric SO(10) unification,''
Phys.\ Rev.\ D {\bf 50}, 7048 (1994).
%[arXiv:hep-ph/9306309].
%%CITATION = HEP-PH 9306309;%% 

\bibitem{logan} 
H.~E.~Logan,
%``Supersymmetric radiative corrections at large tan beta,''
Nucl.\ Phys.\ Proc.\ Suppl.\  {\bf 101}, 279 (2001).
%[arXiv:hep-ph/0102029].
%%CITATION = HEP-PH 0102029;%%

\bibitem{Buchalla:1995vs}
G.~Buchalla, A.~J.~Buras and M.~E.~Lautenbacher,
%``Weak Decays Beyond Leading Logarithms,''
Rev.\ Mod.\ Phys.\  {\bf 68}, 1125 (1996).
%[arXiv:hep-ph/9512380].
%%CITATION = HEP-PH 9512380;%%

\bibitem{Hamzaoui:1998nu}
C.~Hamzaoui, M.~Pospelov and M.~Toharia,
%``Higgs-mediated FCNC in supersymmetric models with large tan(beta),''
Phys.\ Rev.\ D {\bf 59}, 095005 (1999);
%[arXiv:hep-ph/9807350].
%%CITATION = HEP-PH 9807350;%%
K.~S.~Babu and C.~F.~Kolda,
%``Higgs-mediated B0 $\to$ mu+ mu- in minimal supersymmetry,''
Phys.\ Rev.\ Lett.\  {\bf 84}, 228 (2000). 
%[arXiv:hep-ph/9909476].
%%CITATION = HEP-PH 9909476;%%
G.~Isidori and A.~Retico,
%``Scalar flavour-changing neutral currents in the large-tan(beta) limit,''
JHEP {\bf 0111}, 001 (2001).
%[arXiv:hep-ph/0110121].
%%CITATION = HEP-PH 0110121;%%

\bibitem{dedes}
A.~Dedes, H.~K.~Dreiner and U.~Nierste,
%``Correlation of B/s $\to$ mu+ mu- and (g-2)(mu) in minimal supergravity,''
Phys.\ Rev.\ Lett.\  {\bf 87}, 251804 (2001).
%[arXiv:hep-ph/0108037].
%%CITATION = HEP-PH 0108037;%%

\bibitem{bsg}
S. Chen {\it et al.}, CLEO Collaboration,
Phys. Rev. Lett. {\bf 87}, 251807 (2001). 
%BELLE Collaboration, arXiv:hep-ex/0111037.
%To appear in the proceedings of 20th International Symposium
%on Lepton and Photon Interactions at High Energies (Lepton Photon 01),
%Rome, Italy, 23-28 July 2001.
%

\bibitem{arnowitt} 
R.~Arnowitt, B.~Dutta, T.~Kamon and M.~Tanaka,
%``Detection of B/s $\to$ mu+ mu- at the Tevatron run II and constraints on  
% the SUSY parameter space,''
arXiv:hep-ph/0203069.
%%CITATION = HEP-PH 0203069;%%

%\bibitem{Martin:2001st}
%S.~P.~Martin and J.~D.~Wells,
%%``Muon anomalous magnetic dipole moment in supersymmetric theories,''
%Phys.\ Rev.\ D {\bf 64}, 035003 (2001).
%%[arXiv:hep-ph/0103067].
%%%CITATION = HEP-PH 0103067;%%

\bibitem{Feng:1999hg}
J.~L.~Feng and T.~Moroi,
%``Supernatural supersymmetry: 
%Phenomenological implications of  anomaly-mediated supersymmetry breaking,''
Phys.\ Rev.\ D {\bf 61}, 095004 (2000).
%[arXiv:hep-ph/9907319].
%%CITATION = HEP-PH 9907319;%%


\bibitem{bks} S. Baek, P. Ko and W.Y. Song,% KAIST-TH 2002/13, KIAS-P02008,
hep-ph/0208112.

\end{thebibliography}
\end{document}